\documentclass[article,amsmath,11pt,amssymb,floatfix,superscriptaddress,
showpagenumber,showpacs,nofootinbib]{revtex4}
\usepackage{graphicx}
\usepackage{epsfig}
\usepackage{bm}
\usepackage{amsfonts}
\usepackage{xcolor}
\usepackage{caption,subcaption}
\input epsf

\begin{document}
\title{A comprehensive analysis of Barrow holographic Chaplygin gas model reconstruction and its cosmological consequences}

\author{Sanjeeda Sultana}
\email{sanjeeda.sultana0401@gmail.com; sanjeeda.sultana1@s.amity.edu}
\affiliation{ Department of Mathematics, Amity University, Kolkata, Major
Arterial Road, Action Area II, Rajarhat, Newtown, Kolkata 700135,
India.}

\author{Chayan Ranjit}
\email{chayanranjit@gmail.com}
\affiliation{ Department of Mathematics, Egra S. S. B. College, Purba Medinipur 721429, West Bengal, India.}

\author{Surajit Chattopadhyay}
\email{schattopadhyay1@kol.amity.edu; surajitchatto@outlook.com}
\affiliation{ Department of Mathematics, Amity University, Kolkata, Major Arterial Road, Action Area II, Rajarhat, Newtown, Kolkata 700135,
India.(Communicating author)}

\date{\today}

\newpage
\begin{abstract}
 \begin{center}
 \textbf{Abstract}:
 \end{center}
In the current study, we have reconstructed variable modified Chaplygin gas in the Barrow holographic dark energy framework motivated by many recent studies. We have validated the generalized second law of thermodynamics for the reconstructed model. The permissible values of the reconstructed model have been determined by the recent astrophysical and cosmological observational data. The Hubble parameter is presented in terms of the observable parameters and redshift $z$ and other model parameters. From the Stern data set and joint data set of Stern with BAO and CMB observations, the bounds of the model parameters $(B_{0}, \Omega_{bhd0})$ are obtained by the $\chi^{2}$ minimization procedure. The best-fit value of the distance modulus $\mu(z)$ against redshift $z$ is obtained for the reconstructed model and it is consistent with the SNe Ia union2 sample data.\\
\textbf{Keywords:} Barrow holographic dark energy; Variable modified Chaplygin gas; Generalized second law of thermodynamics; Observational data analysis
\end{abstract}
\pacs{98.80.-k; 04.50.Kd}

\maketitle

\section{Introduction} 
A gravitational probe of the universe looks significantly different from an optical one \cite{DE1}. The discrepancy can be categorized and explained by two other components in our Universe: a component without significant dynamics, called dark energy, and a component that acts like cold particles, called cold dark matter. This is made possible by several observational windows on the gravitational side. Both components present significant challenges to our existing knowledge of physics \cite{DE1}. Dark energy (DE) is our main emphasis here (for reviews on DE see \cite{DE1,DE2,DE3,DE4}). In this context it must be mentioned that two teams investigating distant type Ia supernovae (SNeIa) separately provided evidence in 1998 that the current universe is expanding with acceleration \cite{obs1, obs2} and DE is considered to be the driving force behind this accelerated expansion \cite{DE2, DE3, DE4}.

Though theoretical ideas of dark energy date back nearly a century, observable proof has recently emerged in the last 15 years. The ``cosmological constant problem" refers to those theoretical issues \cite{CosmoConst1}. Originating from the holographic principle \cite{hp1, hp2, hp3}, holographic dark energy is an intriguing alternative scenario for the quantitative description of dark energy \cite{hde1,hde2,hde3,hde4,hde5,K1,K2}. Here, we should note that the causality issue may arise in holographic dark energy models. Specifically, the current accelerated expansion may contradict causality since it necessitates the future event horizon to be the universe boundary, which in turn depends on the scale factor's future evolution \cite{hde6}. However, several options have been investigated to deal with this issue. It has been demonstrated that the issue can be resolved by appropriately altering the gravitational sector in the scalar-tensor theories of gravity or by using different modified holographic models, such as Agegraphic dark energy, Ricci dark energy, etc., which use appropriate alternative universe horizon selections \cite{hde6, hde7, hde8, hde9}. In the study by \cite{chatto1}, it was noted that Barrow holographic dark energy (BHDE) \cite{Jawad} reconstructed $f(R)$ gravity allows for a transition in the equation of state from quintessence to phantom. In a particular case of reconstruction, \cite{chatto1} explored the possibility of Little Rip singularity has been observed and the generalized second law of thermodynamics was found to be valid under this reconstruction scheme. 

Incorporating the interaction between holographic dark energy and dark matter allows for the identification of an infrared cutoff at the Hubble radius in a flat universe, which can simultaneously facilitate accelerated expansion and address the coincidence problem. Based on this, \cite{shekhi1} demonstrated that in a non-flat universe the natural choice for the IR cutoff could be the apparent horizon radius. They \cite{shekhi1} showed that any interaction of dark matter with holographic dark energy, whose infrared cutoff is set by the apparent horizon radius, implies an accelerated expansion and a constant ratio of the energy densities of both components thus solving the coincidence problem \cite{shekhi1}. The fulfillment of the generalized second law of thermodynamics in a region enclosed by the apparent horizon has been demonstrated. Their \cite{shekhi1} results hold regardless of the specific form of dark energy. Beginning with the relationship between a quantum field theory's longest length and its ultraviolet cutoff, one can arrive at a holographic vacuum energy, which forms dark energy at cosmological scales. A crucial step in applying the holograpic principle to the cosmological framework is realizing that, like the Bekenstein-Hawking entropy of a black hole, the universe's horizon, or greatest distance, entropy is related to its area \cite{saridakis1}. Just recently, though, Barrow has shown how quantum-gravitational processes could impose complex, fractal characteristics on the black-hole structure, drawing inspiration from the Covid-19 virus images \cite{saridakis1}. Nojiri et al. \cite{hde7} extended the equivalence between the generalized HDE and the Barrow entropic dark energy to the case where the exponent of the Barrow entropy permits to change in response to the universe's cosmological expansion. Nojiri et al. \cite{hde7} calculated the effective EoS parameter from the generalized holographic point of view in both scenarios (whether the Barrow exponent is constant or changes with the cosmological evolution). By comparing with the Barrow DE EoS parameter, this further guarantees the equivalence between the generalized HDE and the Barrow entropic dark energy \cite{hde7}. In a recent work, Oliveros et al. \cite{hde8} examined the cosmic evolution of the Granda–Oliveros infrared cutoff and its impact on the recently adopted BHDE model. They \cite{hde8} explained how the evolution of $H(z)$ is influenced by the deformation parameter $\Delta$ and that an accelerated expansion regime of the universe at late times can be obtained from this model. 

Motivated by the fact that the Chaplygin gas has a negative pressure, researchers have taken on the straightforward task of studying a FRW cosmology of a universe filled with this kind of fluid \cite{setare1}. Among the various candidates to play the role of the dark energy, the Chaplygin gas has emerged as a potential unification of dark matter and dark energy because its cosmological evolution is similar to an initial dust-like matter and a cosmological constant for late times. \textcolor{black}{In the current work, we have associated the BHDE in FRW universe with a Chaplygin cosmology. This kind of consideration of connectivity between Chaplygin gas and candidates of holographic dark energy is not new in the literature. A very notable work towards the reconstruction approach is due to authors \cite{VS}. In the study of Setare \cite{setare1} we find that they considered a correspondence between the holographic dark energy density and Chaplygin gas energy density in FRW universe and reconstructed the potential and the dynamics of the scalar field which describe the Chaplygin cosmology. Considering correspondence between holographic dark energy density and the interacting generalized Chaplygin gas energy density in the FRW universe \cite{setare2}.  In a different study, \cite{chattopadhyay1} examined the interaction scenario of a universe that contained both regular matter and variable modified Chaplygin gas. The holographic dark energy density and the interaction variable modified Chaplygin gas energy density were thought to coincide, according to the authors of \cite{chattopadhyay1}. We have seen a reconstruction scheme by considering a correspondence between the tachyon DE model and BHDE in a notable work \cite{R1}. The latter is a modified situation where the holographic principle is applied using Barrow entropy rather than the standard Bekenstein–Hawking one. The study in \cite{R2} explores a reconstruction scheme for the k-essence form of dark energy with the most generalized version of HDE proposed in \cite{NO,NO1}. In another noteworthy work \cite{L1}, the authors have considered a cosmological model in the framework of Einstein–Cartan theory with a
single scalar torsion $\phi=\phi(t)$ and reconstructed the torsion model corresponding to the HDE density.}

In this work, we have explored the generalized second law (GSL) in the cosmological context and it can be interpreted that the time derivative of generalized entropy which is the sum of the entropy of the cosmological horizon and the entropy of all fluids filling the space must be increasing or non-decreasing function of time. According to Setare \cite{G1}, the GSL is respected for specific deceleration parameter values in a universe dominated by holographic dark energy. Additionally, certain lower-dimensional cosmological scenarios have been used to study the GSL, with some intriguing results \cite{G2}. The cosmological scenario where dark energy interacts with both dark matter and radiation \cite{G3} has been used to examine the validity of GSL of thermodynamics. Regardless of the background geometry and the particular interaction form of the fluids equation of state, it has been demonstrated that the GSL is always and generally true. Furthermore, the GSL of thermodynamics \cite{G4,G5} and viscous dark energy and thermodynamics of viscous dark energy in the Rundall-Sundram II \cite{G6} braneworld \cite{G4,G5} have been studied.  Gong et al. \cite{G7} have addressed the thermodynamics of DE by considering the DE models with constant $w$ and the generalized Chaplygin gas (GCG). In \cite{G8}, the validity of GSL of thermodynamics is studied for holographic dark energy interacting with two fluids. 

Given all these, we have explored a correspondence between variable modified Chaplygin gas and Barrow holographic dark energy framework in the current study. The rest of the paper is organized as follows: In Section II, we have reconstructed variable modified Chaplygin gas in the framework of Barrow holographic dark energy. In Section III, we have shown the validity of the GSL of thermodynamics for the reconstructed model. In Section IV, we have demonstrated the computational basis of the analysis of the reconstructed model with the
observational data. In Section V, we have analyzed observational data. In Section VI, we have concluded.

\section{Barrow holographic reconstruction of variable modified Chaplygin gas}

In this section, we will construct the scenario of VMCG in the framework of BHDE. \textcolor{black}{Considering the reconstruction endeavour in mind, let us now mention some aspects of BHDE. The BHDE is effectively a generalization of standard HDE, where a new free parameter termed as deformation parameter $\Delta$ can impact the results of the models involving BHDE. Saridakis \cite{saridakis1} elaborately demonstrated the scenario of BHDE by applying the conventional holographic principle in the cosmological framework. It has been demonstrated in \cite{saridakis1} that for the BHDE scenario, the Barrow entropy is utilized instead of Bekenstein-Hawking entropy. In this connection it may be noted, that Barrow proposed that quantum gravitational effects may bring into the fractal structure on the surface of black hole, thereby leading to deformed entropy, quantified by $\Delta$. Therefore BHDE gives rise to novel cosmological scenarios for $\Delta>0$ and up to the maximal deformation for $\Delta=1$ \cite{saridakis1,Barrow}.} In the literature, the analysis of the connection between Chaplygin gas and candidates of holographic dark energy is not new. A correspondence between the holographic dark energy density and Chaplygin gas energy density in FRW universe has been considered in the study of Setare \cite{setare1} and the potential and the dynamics of the scalar field which describe the Chaplygin cosmology are reconstructed. In another study, we have seen correspondence between the interacting generalized Chaplygin gas energy density and holographic dark energy density in the FRW universe \cite{setare2}. The interaction scenario of a universe that contained both regular matter and variable modified Chaplygin gas has been examined in a different study \cite{chattopadhyay1}. According to the authors of \cite{chattopadhyay1}, the holographic dark energy density and the interaction variable modified Chaplygin gas energy density were thought to coincide. Given these, we proceed to the following reconstruction model.

\subsection{The reconstruction scheme}

In an FRW model, the metric of a spatially flat homogeneous and isotropic universe is
\begin{equation}
ds^{2}=-dt^{2}+a^{2}(t)[dr^{2} +r^{2}(d\theta^{2}+sin^{2}\theta d\phi^{2})].
\label{E1}
\end{equation}
where $a(t)$ is the scale factor of the universe and $(t,r,\theta,\phi)$ are the
comoving coordinates. In the current study, we have used redshift $z$ in various phases, given by $z=a^{-1}-1$ \cite{1}. The Friedmann equation and the acceleration equation are
\begin{equation}
H^{2}=\frac{1}{3}(\rho_{DM}+\rho_{DE})
\label{E2}
\end{equation}
and 
\begin{equation}
\dot{H}=-\frac{1}{2}(\rho_{DM}+\rho_{DE}+p_{DE})
\label{E3}
\end{equation}
respectively, where $H$ is the Hubble parameter which expresses the Universe's expansion rate and has units of inverse time or mass units in natural units. It is defined by $H \equiv \frac{\dot{a}}{a}$ ($\dot{a}=\frac{da}{dt}$). $p_{DE}$ and $\rho_{DE}$ are isotropic pressure and energy density for dark energy respectively, and $\rho_{DM}$ is the energy density for dark matter. Here we have chosen $8\pi G = c = 1$. As the dark matter is considered to be pressureless, hence $p_{DM}$ is equivalent to $0$. The conservation equation is given by
\begin{equation}
    \dot{\rho}_{total}+3H(\rho_{total}+p_{total})=0,
    \label{E4}
\end{equation}
where $\rho_{total}=\rho_{DE}+\rho_{DM}$ and $p_{total}=p_{DE}$ (as $p_{DM}=0$). The equation of state for VMCG \cite{2} is
\begin{equation}
    p_{VMCG}=A\rho_{VMCG}-\frac{B(a)}{\rho_{VMCG}^{\alpha}}.
    \label{E5}
\end{equation}
The inequality $\rho_{DE} L^{4} \leq S$ corresponds to the  standard HDE, where $L$ is the horizon length, and the use of Barrow entropy \cite{4} under the imposition $S \propto A \propto L^{2}$ \cite{5} leads the form of the energy density of BHDE \cite{3} as
\begin{equation}
\rho_{BHDE}=\gamma L^{\Delta-2},
    \label{E6}
\end{equation}
where $\gamma$ is a parameter whose dimension is $[L]^{-2-\Delta}$. In the case of a spatially flat universe, $\kappa=0$. If the enveloping horizon is considered to be the apparent horizon then $L=\frac{1}{\sqrt{H^{2}+\frac{\kappa}{a^{2}}}}$. Hence for a flat universe, $L=\frac{1}{H}$.The abovementioned expression provides the standard HDE $\rho_{DE}=3 c^{2} M_{p}^{2} L^{-2}$ in the scenario where $\Delta=0$. Here $M_{p}$ is the Planck mass and $C=3 c^{2} M_{p}^{2}$ where $c$ is the model parameter. Assuming $B(a)=B_{0}a^{-n}$ in Eq.(\ref{E5}) with constants  $B_{0}>0$ and $n>0$ we have the form of EOS parameter ($w=\frac{p}{\rho}$) for VMCG as
\begin{equation}
w_{VMCG}=A-\frac{B_{0}a^{-n}}{\rho_{VMCG}^{\alpha+1}}.
    \label{E7}
\end{equation}
By using Eq.(\ref{E7}) in Eq.(\ref{E4}), we have the following ordinary differential equation
\begin{equation}
\frac{d\rho_{VMCG}}{da} +\frac{3}{a} \rho_{VMCG} \biggl(1+A-\frac{B_{0}a^{-n}}{\rho_{VMCG}^{\alpha+1}}\biggr)=0,
    \label{E8}
\end{equation}
while solving Eq.(\ref{E8}), we got the reconstructed density for VMCG as
\begin{equation}
\rho_{VMCG}=\left(\frac{3 a^{-n} B_{0} (1+\alpha )}{3-n+3 \alpha +3 A (1+\alpha )}+a^{-3 (1+A) (1+\alpha )} C_{1}\right)^{\frac{1}{1+\alpha
}},
    \label{E9}
\end{equation}
where $C_{1}$ is the integration constant which is denoted as $\rho_{VMCG0}$ in the subsequent expressions. On the other hand, BHDE will deviate from the standard one, resulting in other cosmological behaviour, under the scenario when the distortion effects measured by $\Delta$ turn on. It is now possible for us to determine the relationship between BHDE and VMCG. To do this, we take the effective underlying theory to be the BHDE. To apply the correspondence between BHDE and VMCG, we determine $\rho_{VMCG}=\rho_{BHDE}$. This equality can be taken into account since Eq.(\ref{E2}) allows us to establish dimensionless critical densities $\Omega_{DE}=\frac{\rho_{DE}}{3 H^{2}}$, and we are utilizing the BHDE as the effective underlying theory. The references \cite{6,7,8,9} have previously used this kind of strategy. While taking into account the correspondence between $\rho_{VMCG}$ and $\rho_{BHDE}$ with the Hubble horizon as the enveloping horizon we get the reconstructed Hubble parameter as
\begin{equation}
H=\left(\frac{\left(a^{-3 (1+A) (1+\alpha )} \rho_{VMCG0}+\frac{3 a^{-n} B_{0} (1+\alpha )}{3-n+3 \alpha +3 A (1+\alpha )}\right)^{\frac{1}{1+\alpha
}}}{\gamma}\right)^{\frac{1}{2-\Delta }}.
\label{E10}
\end{equation}
By using the relation $\dot{H}=\frac{a}{2}\frac{dH^{2}}{da}$ and Eq.(\ref{E2}) in the Eq.(\ref{E3}), we have the expression for the reconstructed EoS parameter for the Barrow holographic variable modified Chaplygin gas (BHVMCG) model as
\begin{equation}
\begin{array}{cc}
   w_{reconstruct}=\frac{-a^{3 (1+A) (1+\alpha )} B_{0} (2 n+3 (1+\alpha ) (-2+\Delta ))+a^n \rho_{VMCG0} (-3+n-3 \alpha -3 A (1+\alpha )) (2 A+\Delta
)}{\left(3 a^{3 (1+A) (1+\alpha )} B_{0} (1+\alpha )+a^n \rho_{VMCG0} (3-n+3 \alpha +3 A (1+\alpha ))\right) (-2+\Delta )}.
\end{array}
    \label{E11}
\end{equation}
\begin{figure}
\begin{center}
\includegraphics[height=3.0in]{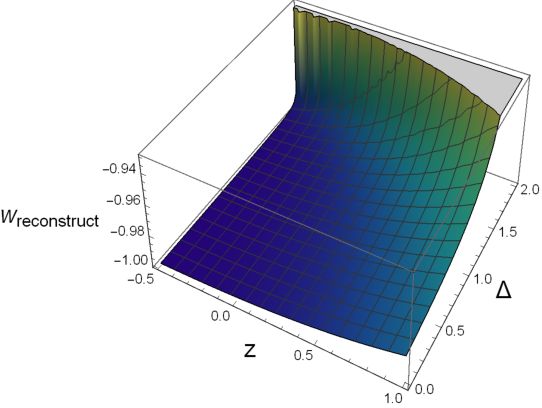}
\caption{\textcolor{black}{Evolution of reconstructed EoS parameter $w_{reconstruct}$ against redshift $z$ and $\Delta$ for the BHVMCG model. The parameters chosen are $A=0.075$, $\alpha=0.004$, $n=0.00001$, $B_{0}=4$, and $\rho_{VMCG0}=0.006$.}}
\label{F1}
\end{center}
\end{figure}

\textcolor{black}{In Fig.\ref{F1}, we have plotted the reconstructed EoS parameter $w_{reconstruct}$ against redshift $z$ and $\Delta$ for the BHVMCG model based on Eq.(\ref{E11}). We infer from Fig.\ref{F1} that it shows a quintessence behaviour i.e. $-1<w_{reconstruct}<-\frac{1}{3}$, and the reconstructed EoS parameter is exhibiting a decaying pattern with the evolution of the universe. In this pictorial presentation, we have varied both $z$ and $\Delta$ and observed significant variability of the EoS parameter with both of them. Irrespective of the values of $\Delta$ close to $0$ or $2$, the decreasing pattern of the EoS parameter is always there. However, it is notable that for higher values of $\Delta$, the EoS parameter is decreasing more sharply than values of $\Delta$ close to $0$. However, despite this change in the decaying pattern, the value of the reconstructed EoS parameter is close to $-1$ for the current universe i.e. at $z=0$. In all the cases, we have seen that in the later stage, it becomes asymptotic in the neighbourhood of phantom boundary $-1$. Thus we understand that the reconstructed EoS parameter behaves like quintessence for the range of the values of the deformation parameter between $0$ and $2$. It is noteworthy that at $z=0$, $w_{reconstruct} \approx -1$ i.e. it behaves as a cosmological constant.}

By substituting Eq.(\ref{E11}) in Eq.(\ref{E7}), we have the energy density for the BHVMCG model as
\begin{equation}
\begin{array}{c}
\rho_{reconstruct}=\left(\left(a^n \left(a^{3 (1+A) (1+\alpha )} B_{0} (2 n+3 (1+\alpha ) (-2+\Delta )+3 A (1+\alpha ) (-2+\Delta ))+a^n
(1+A) \rho_{VMCG0}\right.\right.\right.\\
\left.\left.\left. (3-n+3 \alpha +3 A (1+\alpha )) \Delta \right)\right)
\left(B_{0} \left(3 a^{3 (1+A) (1+\alpha )} B_{0} (1+\alpha )+a^n \rho_{VMCG0} (3-n+3
\alpha +3 A (1+\alpha ))\right) \right.\right.\\
\left.\left.(-2+\Delta )\right)\right)^{-\frac{1}{1+\alpha }}.
\end{array}
    \label{E12}
\end{equation}
From the relation $w=\frac{p}{\rho}$, we have the pressure for the BHVMCG model as
\begin{equation}
\begin{array}{c}
p_{reconstruct}=\left(\left(\frac{a^n \left(a^{3 (1+A) (1+\alpha )} B_{0} (2 n+3 (1+\alpha ) (-2+\Delta )+3 A (1+\alpha ) (-2+\Delta ))+a^n
(1+A) \rho_{VMCG0} (3-n+3 \alpha +3 A (1+\alpha )) \Delta \right)}{B_{0} \left(3 a^{3 (1+A) (1+\alpha )} B_{0} (1+\alpha )+a^n \rho_{VMCG0} (3-n+3
\alpha +3 A (1+\alpha ))\right) (-2+\Delta )}\right)^{-\frac{1}{1+\alpha }}\right.\\ 
\left(-a^{3 (1+A) (1+\alpha )} B_{0} (2 n+3 (1+\alpha ) (-2+\Delta
))+a^n \rho_{VMCG0} (-3+n-3 \alpha -3 A (1+\alpha )) (2 A+\Delta )\right)\\
\left(\left(3 a^{3 (1+A) (1+\alpha )} B_{0} (1+\alpha )+a^n \rho_{VMCG0} (3-n+3
\alpha +3 A (1+\alpha ))\right) (-2+\Delta )\right)^{-1}.
\end{array}
     \label{E13}
\end{equation}
From Eq.\ref{E2}, we have the matter density for the BHVMCG model as 
\begin{equation}
\begin{array}{c}
\rho_{DM,reconstruct}=3 \left(\frac{\left(a^{-3 (1+A) (1+\alpha )} \rho_{VMCG0}+\frac{3 a^{-n} B_{0} (1+\alpha )}{3-n+3 \alpha +3 A (1+\alpha
)}\right)^{\frac{1}{1+\alpha }}}{\gamma }\right)^{-\frac{2}{-2+\Delta }}-\\
\left(\frac{a^n \left(a^{3 (1+A) (1+\alpha )} B_{0} (2 n+3 (1+\alpha ) (-2+\Delta
)+3 A (1+\alpha ) (-2+\Delta ))+a^n (1+A) \rho_{VMCG0} (3-n+3 \alpha +3 A (1+\alpha )) \Delta \right)}{B_{0} \left(3 a^{3 (1+A) (1+\alpha )} B_{0}
(1+\alpha )+a^n \rho_{VMCG0} (3-n+3 \alpha +3 A (1+\alpha ))\right) (-2+\Delta )}\right)^{-\frac{1}{1+\alpha }}.
\end{array}
    \label{E14}
\end{equation}
Hence, the reconstructed total EoS parameter for the BHVMCG model is
\begin{equation}
\begin{array}{c}
w_{total,reconstruct}=\left(\left(\frac{\left(a^{-3 (1+A) (1+\alpha )} \rho_{VMCG0}+\frac{3 a^{-n} B_{0} (1+\alpha )}{3-n+3 \alpha +3 A (1+\alpha )}\right)^{\frac{1}{1+\alpha
}}}{\gamma}\right)^{\frac{2}{-2+\Delta }}\right.\\
 \left(\frac{a^n \left(a^{3 (1+A) (1+\alpha )} B_{0} (2 n+3 (1+\alpha ) (-2+\Delta )+3 A (1+\alpha ) (-2+\Delta
))+a^n (1+A) \rho_{VMCG0} (3-n+3 \alpha +3 A (1+\alpha )) \Delta \right)}{B_{0} \left(3 a^{3 (1+A) (1+\alpha )} B_{0} (1+\alpha )+a^n \rho_{VMCG0}
(3-n+3 \alpha +3 A (1+\alpha ))\right) (-2+\Delta )}\right)^{-\frac{1}{1+\alpha }} \\
\left.\left(-a^{3 (1+A) (1+\alpha )} B_{0} (2 n+3 (1+\alpha ) (-2+\Delta ))+a^n \rho_{VMCG0} (-3+n-3 \alpha -3 A (1+\alpha )) (2 A+\Delta )\right)\right.\\
\left(3
\left(3 a^{3 (1+A) (1+\alpha )} B_{0} (1+\alpha )+a^n \rho_{VMCG0} (3-n+3 \alpha +3 A (1+\alpha ))\right) (-2+\Delta )\right)^{-1}.
\end{array}
\label{E15}
\end{equation}
\begin{figure}
\begin{center}
\includegraphics[height=3.0in]{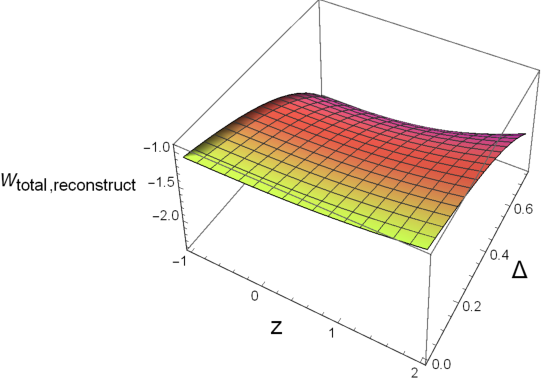}
\caption{\textcolor{black}{Evolution of reconstructed total EoS parameter $w_{total,reconstruct}$ against redshift $z$ and $\Delta$ for the BHVMCG model. The parameters chosen are $A=0.003$, $\alpha=0.004$, $n=0.001$, $B_{0}=0.9$, $\rho_{VMCG0}=0.006$ and $\gamma=3.3$.}}
\label{F2}
\end{center}
\end{figure}
\textcolor{black}{In Fig.\ref{F2}, we have plotted the reconstructed total EoS parameter $w_{total,reconstruct}$ against redshift $z$ and $\Delta$ for the BHVMCG model based on Eq.(\ref{E15}). We infer from Fig.\ref{F2} that the behaviour of this EoS parameter has some significant variation with the values of the deformation parameter $\Delta$. For the values of $\Delta$ close to $0$, the EoS parameter is asymptotic in the neighbourhood of $-1$ and for such values of $\Delta$ the behaviour of the EoS parameter is quintessence. However, a significant change becomes visible as $\Delta$ goes away from $0$. Nearly from $\Delta \approxeq 0.1$, the EoS parameter starts exhibiting a quintom behaviour i.e. a transition from $w_{total,reconstruct}>-1$ to $w_{total,reconstruct}<-1$ which means there is a crossing of phantom boundary and is showing a monotone decreasing pattern with the evolution of the universe. Hence, for values of $\Delta \gtrapprox 0.1$, the $w_{total,reconstruct}$ exhibits a transition from quintessence to phantom, in contrast to the $w_{reconstruct}$, where we have witnessed quintessence behaviour. Furthermore, it should be highlighted that there is no future escape from phantom even when the $w_{total,reconstruct}$ transits to quintom in the later stage of the Universe. Therefore, it does not suggest that big-rip singularity can be avoided for the BHVMCG model.}

Now let's talk about the fractional densities. To do that, we first take into account the critical density, $\rho_{c} = 3H^{2}$ (we have already chosen $8\pi G = 1$). Thus, the following are the fractional densities for dark matter and dark energy:
\begin{equation}
\Omega_{DE}=\frac{\rho_{DE}}{3 H^{2}},
    \label{E16}
\end{equation}
\begin{equation}
\Omega_{DM}=\frac{\rho_{DM}}{3 H^{2}}.
    \label{E17}
\end{equation}
Consequently, we can write 
\begin{equation}
\Omega_{DM}+\Omega_{DE}=1.
    \label{E18}
\end{equation}
Using Eqs. (\ref{E10}) and (\ref{E12}) in Eq.(E16), we have the reconstructed fractional density for the BHVMCG model as
\begin{equation}
\begin{array}{c}
\Omega_{reconstruct}=\frac{1}{3} \left(\frac{\left(a^{-3 (1+A) (1+\alpha )} \rho_{VMCG0}+\frac{3 a^{-n} B_{0} (1+\alpha )}{3-n+3 \alpha +3 A (1+\alpha
)}\right)^{\frac{1}{1+\alpha }}}{\gamma}\right)^{\frac{2}{-2+\Delta }}\\
 \left(\frac{a^n \left(a^{3 (1+A) (1+\alpha )} B_{0} (2 n+3 (1+\alpha ) (-2+\Delta
)+3 A (1+\alpha ) (-2+\Delta ))+a^n (1+A) \rho_{VMCG0} (3-n+3 \alpha +3 A (1+\alpha )) \Delta \right)}{B_{0} \left(3 a^{3 (1+A) (1+\alpha )} B_{0}
(1+\alpha )+a^n \rho_{VMCG0} (3-n+3 \alpha +3 A (1+\alpha ))\right) (-2+\Delta )}\right)^{-\frac{1}{1+\alpha }}.
\end{array}
    \label{E19}
\end{equation}
The evolution of the deceleration parameter $q$ \cite{10,11} against the redshift $z$ is given by
\begin{equation}
q=-\frac{a\ddot{a}}{\dot{a}^{2}}=-1-\frac{\dot{H}}{H^{2}}.
    \label{E20}
\end{equation}
Hence we get the the reconstructed deceleration parameter for the BHVMCG model as
\begin{equation}
\begin{array}{c}
q_{reconstruct}=\frac{-3 a^{3 (1+A) (1+\alpha )} B_{0} (n+(1+\alpha ) (-2+\Delta ))+a^n \rho_{VMCG0} (-3+n-3 \alpha -3 A (1+\alpha )) (1+3 A+\Delta)}{\left(3 a^{3 (1+A) (1+\alpha )} B_{0} (1+\alpha )+a^n \rho_{VMCG0} (3-n+3 \alpha +3 A (1+\alpha ))\right) (-2+\Delta )}.
\end{array}
\label{E21}
\end{equation}
\begin{figure}
\begin{center}
\includegraphics[height=3.0in]{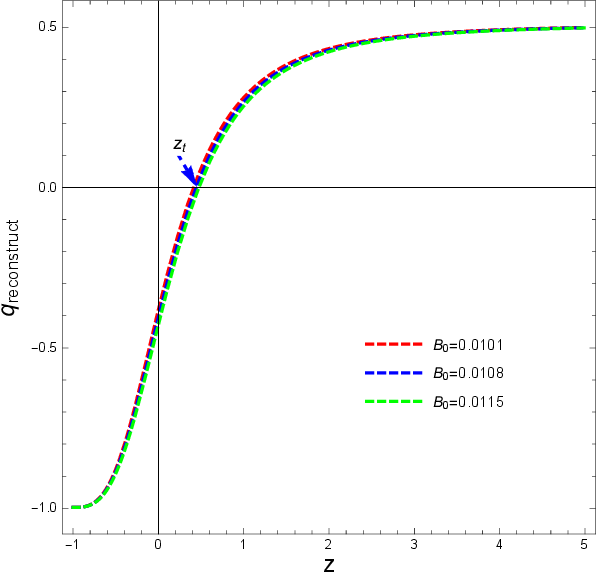}
\caption{Evolution of reconstructed deceleration parameter $q_{reconstruct}$ for the BHVMCG model against redshift $z$. The parameters chosen are $A=0.002$, $\alpha=0.004$, $n=0.005$, $\rho_{VMCG0}=0.007$ and $\Delta=0.008$.}
\label{F3}
\end{center}
\end{figure}
In Fig.\ref{F3}, we have plotted the reconstructed deceleration parameter $q_{reconstruct}$ against redshift $z$ for the BHVMCG model based on Eq.(\ref{E21}). From Fig.\ref{F3}, we observe that at the very early stage of the universe, $q_{reconstruct} > 0$, roughly around $z > 0.5$,  i.e. the decelerated expansion phase of the universe. A transition is seen in the case of the deceleration parameter $q_{reconstruct}$ at $z \approx 0.5$  from a positive to a negative region, which means the universe gradually transits from the decelerated $(q_{reconstruct} > 0)$ to the accelerated expansion phase $(q_{reconstruct} < 0)$ and the transition occurs at $z_{t} \approx 0.45$ which is consistent with the observations in the literature \cite{12}. In the current stage of the Universe it converges towards -$1$ and becomes asymptotic in its neighborhood of -$1$. Hence, we infer that a transition is possible from the decelerated expansion phase to the accelerated expansion phase of the universe in the case of the BHVMCG model. 

\subsection{Statefinder and $O_{m}$ diagnostic}
The cosmological diagnostic pair $(r,s)$ \cite{13,14} is directly dependent on the scale factor $a$ and hence on the metric describing space-time, that's why it is regarded as “geometrical”. The parameters $r$ and $s$ in terms of Hubble parameter $H$ and its time derivatives as
\begin{equation}
r=1+3\frac{\dot{H}}{H^{2}}+\frac{\ddot{H}}{H^{3}}
    \label{E22}
\end{equation}
and
\begin{equation}
s=-\frac{3H\dot{H}+\ddot{H}}{3H(2\dot{H}+3H^{2})}
    \label{E23}
\end{equation}
respectively. Different combinations of $r$ and $s$ are represented by different DE models \cite{13,14} as follows:
\begin{itemize}
\item $(r=1,s=0)$ corresponds $\Lambda$CDM model
\item $(r=1,s=\frac{2}{3})$ corresponds HDE model
\item $(r=1,s=1)$ corresponds SCDM model
\item $(r>1,s<0)$ corresponds CG model
\item $(r<1,s>0)$ corresponds Quintessence region
\end{itemize}
The evolutionary trajectories of the HDE model in the $s-r$ plane \cite{15,16,17,18,19,20,21} begins at $s=\frac{2}{3},r=1$ with the future event horizon as the IR cut-off and in due course approach the $\Lambda$CDM fixed point $(s = 0, r = 1)$. According to \cite{22}, the $s-r$ plane curves are vertical for both the quintessence DE model with a constant EoS parameter \cite{13,14} and the Ricci DE (RDE) model.  In Chaplygin gas (CG), The $s-r$ plane trajectory for the CG model lies in the regions $s < 0, r > 1$ \cite{23}. On the other hand, both the quintessence (inverse power-law) models $(Q)$ and the phantom model with power law potential lie in the regions $s > 0, r < 1$ \cite{13,14}, and both approach the $\Lambda$CDM fixed point at late time. The trajectory in the $s-r$ plane creates a swirl in the coupled quintessence models \cite{24} before reaching the attractor. The  Polytropic gas model \cite{25} and the Agegraphic DE model \cite{26} both exhibit $\Lambda$CDM behaviour at the early universe epoch. In the $(s, r)$ plane \cite{27}, similar behaviour is shown by the HDE model of DE with model parameter $c = 1$ and the ghost DE model. Models of DE such as HDE \cite{19,20,21}, Yang–Mills \cite{28}, generalized Chaplygin gas \cite{29,30,31}, new agegraphic \cite{26,32} and Chaplygin gas \cite{33,34} is consistent with this behaviour. In the case of the HDE model with Granda–Oliveros IR cut-off\cite{35} and the tachyon DE model \cite{36}, the $s-r$ plane curve passes through the $\Lambda$CDM fixed point at the middle of the evolution of the universe. At late time the $s-r$ plane trajectories terminate at the $\Lambda$CDM fixed point $(s = 0, r = 1)$ in the case of Tsallis HDE. They traverse an arc segment and a parabola (downward) \cite{37,38} after starting at the matter-dominated Standard Cold Dark Matter (SCDM) $s = 1, r = 1$. The evolutionary curve in the $s-r$ plane represents the Chaplygin gas behaviour in the case of a Ricci HDE model \cite{39,40}. It starts and ends with a swirl at the $Lambda$CDM fixed point $(s = 0, r = 1)$. According to recent work by one of the authors\cite{41}, the Statefinder pair $(r,s)$ of the SMHDE model always lies in the Chaplygin gas region and approaches the $\Lambda$CDM fixed point $(r = 1, s = 0)$ in the late time evolution of the universe. \cite{36} examines the evolution of the $(r,s)$ pair for the new generalized CG model. The $s-r$ plane's evolutionary curve starts at a cosmological constant, turns a corner, and travels to a distinct endpoint in the scenario of the Tsallis agegraphic dark energy model \cite{43}. Conversely, the Statefinder pair analysis for a range of DE models has been extensively explored by the authors of \cite{44,45}. 

Using the expression for the Hubble parameter and its time derivatives, we have the reconstructed parameters $r$ and $s$ as
\begin{equation}
\begin{array}{c}
r_{reconstruct}=\left(9 a^{-2 n} B_{0}^2 (n+(1+\alpha ) (-2+\Delta )) (2 n+(1+\alpha ) (-2+\Delta ))+\right.\\
a^{-6 (1+A) (1+\alpha )} \rho_{VMCG0}^2 (3-n+3 \alpha
+3 A (1+\alpha ))^2 (1+3 A+\Delta ) (4+6 A+\Delta )-\\
3 a^{-n-3 (1+A) (1+\alpha )} B_{0} \rho_{VMCG0} (3-n+3 \alpha +3 A (1+\alpha )) \\
\left.\left.\left(-6 A (3+\alpha  (9-2 n+6 \alpha ))+n (6-6 \alpha  (-2+\Delta )-9 \Delta )+n^2 (-2+\Delta )+9 A^2 (1+\alpha )^2 (-2+\Delta)+\right.\right.\right.\\
\left.\left.(1+\alpha ) (4+9 \alpha -2 \Delta ) (-2+\Delta )+3 A (1+\alpha ) (3-2 n+6 \alpha ) \Delta \right)\right)\\
\left(\left(3 a^{-n} B_{0} (1+\alpha )+a^{-3 (1+A) (1+\alpha )} \rho_{VMCG0} (3-n+3 \alpha +3 A (1+\alpha ))\right)^2 (-2+\Delta )^2\right)^{-1},
\end{array}
\label{E24}
\end{equation}
and 
\begin{equation}
\begin{array}{c}
s_{reconstruct}=\left(-3 a^{6 (1+A) (1+\alpha )} B_{0}^2 n (2 n+3 (1+\alpha ) (-2+\Delta ))-\right.\\
3 a^{2 n} (1+A) \rho_{VMCG0}^2 (3-n+3 \alpha +3 A (1+\alpha
))^2 \\
(2 A+\Delta )+a^{n+3 (1+A) (1+\alpha )} B_{0} \rho_{VMCG0} (3-n+3 \alpha +3 A (1+\alpha )) \\
\left.\left.\left(-6 A (3+\alpha  (9-2 n+6 \alpha ))+n (6-6 \alpha  (-2+\Delta )-9 \Delta )+n^2 (-2+\Delta )+
9 \alpha  (1+\alpha ) (-2+\Delta)+\right.\right.\right.\\
\left.\left.9 A^2 (1+\alpha )^2 (-2+\Delta )+3 A (1+\alpha ) (3-2 n+6 \alpha ) \Delta \right)\right)\\
\left(3 \left(3 a^{3 (1+A) (1+\alpha )} B_{0} (1+\alpha )+a^n \rho_{VMCG0} (3-n+3 \alpha +3 A (1+\alpha ))\right) (-2+\Delta )\right.\\
\left. \left(a^{3 (1+A)(1+\alpha )} B_{0} (2 n+3 (1+\alpha ) (-2+\Delta ))+a^n \rho_{VMCG0} (3-n+3 \alpha +3 A (1+\alpha )) (2 A+\Delta )\right)\right)^{-1}
\end{array}
\label{E25}
\end{equation}
respectively.
\begin{figure}
\begin{center}
\includegraphics[height=3.5in]{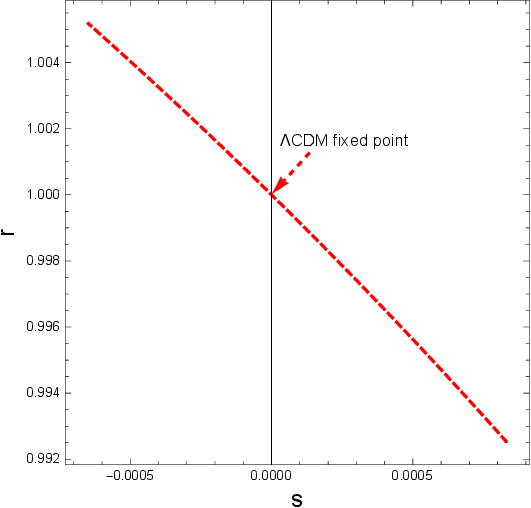}
\caption{The statefinder pair $(r,s)$ trajectory for the BHVMCG model. The parameters chosen are $A=0.02$, $B_{0} = 0.001$, $\alpha=0.004$, $\rho_{VMCG0}=0.003$, $n=0.005$ and $\Delta=0.001$.}
\label{F4}
\end{center}
\end{figure}

\textcolor{black}{The evolution of the reconstructed Statefinder pair $(r,s)$ for the BHVMCG model has been plotted in Fig.\ref{F4} based on Eqs. (\ref{E24}) and (\ref{E25}). From Fig.\ref{F4}, we can infer that the evolutionary trajectory of the Statefinder pair for the model starts its evolution from the region $(s>0,r<1)$ and as time passes it traverses through the $\Lambda$CDM fixed point $(s=0,r=1)$. As already described, the statefinder trajectory traverses the quintessence phase i.e. $(s>0,r<1)$  to reach $\Lambda$CDM fixed point. At the later stage, it lies in the Chaplygin gas region i.e. $(s<0,r>1)$. As the trajectory passes through the $\Lambda$CDM fixed point, it confirms strongly that the BHVMCG model circulates the $\Lambda$CDM phase of the universe. Moreover, from Fig.\ref{F4}, it is also understandable that $s$ has a tendency to go to $-\infty$ with finite $r$. Thus the possibility of interpolation between universe's dust and $\Lambda$CDM phase is also indicated here.}

To compare DE models with the $\Lambda$CDM model \cite{46,47}, we have another diagnostic namely $O_{m}(z)$. The Hubble parameter $H$ and redshift $z$ yield this parameter which is defined as
\begin{equation}
O_{m}(z)=\frac{E(z)-1}{(1+z)^{3}-1},
    \label{E26}
\end{equation}
where $E(z)=\frac{H(z)}{H_{0}}$ is a dimensionless parameter for the rate of expansion of the universe and $H_{0}$ denotes the current value of the Hubble parameter $H$. Based on this diagnostic, the following behaviours are exhibited by DE: zero curvature corresponds to $O_{m}(z)=\Lambda$CDM, positive curvature corresponds to phantom type behaviour, and negative curvature corresponds to quintessence type behaviour. The slope of $O_{m}(z)$ can distinguish variations in DE models despite inexplicit knowledge of the dark matter density. 

\begin{figure}
\begin{center}
\includegraphics[height=3.5in]{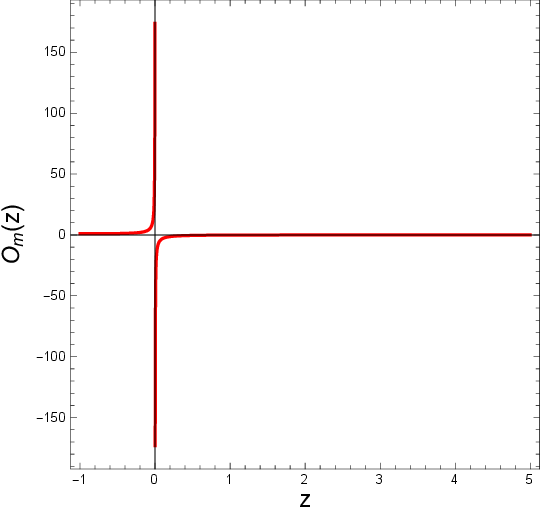}
\caption{Evolution of reconstructed $O_{m}(z)$ diagnostic with respect to redshift $z$ for the BHVMCG model. The parameters chosen are $A = 0.002$, $\gamma = 0.02$, $B_{0}=0.02$, $\alpha=0.5$, $\delta =0.002$,  $\rho_{VMCG0}=0.4$, $n=0.001$, $h_{0}=73.8$ and $\beta=0.4$.}
\label{F5}
\end{center}
\end{figure}

Evolution in the $O_{m}(z)$ diagnostic value indicates a departure from the $\Lambda$CDM model. Given that the $O_{m}(z)$ diagnostic solely depends on the expansion rate, it can be easily ascertained using the current observations. From the reconstructed Hubble parameter, we have the 
$O_{m}(z)$ diagnostic for the BHVMCG model as
\begin{equation}
O_{m}(z)=\frac{-1}{-1+(1+z)^3}+\frac{\left(\frac{\left(a^{-3 (1+A) (1+\alpha )} \rho_{VMCG0}+\frac{3 a^{-n} B_{0} (1+\alpha )}{3+3 A-n+3 (1+A) \alpha
}\right)^{\frac{1}{1+\alpha }}}{\gamma}\right)^{-\frac{2}{-2+\Delta }}}{h_{0}^{2}(-1+(1+z)^3)}.
    \label{E27}
\end{equation}
The reconstructed $O_{m}(z)$ diagnostic has been plotted against redshift $z$ for the BHVMCG model based on Eq.(\ref{E27}) in Fig.\ref{F5}. We know that the negative curvature of $O_{m}(z)$ trajectories displays the quintessence behaviour of DE, whereas the phantom behaviour corresponds to its positive curvature. In Fig.\ref{F5} both the regions have been shown by the $O_{m}(z)$ trajectories.

\section{Thermodynamic analysis}
The cosmological application of holography \cite{48} relies heavily on the fact that, like a black hole, the entropy of the entire universe, when seen as a system with a radius equal to the previously stated maximum distance, is proportional to its area. The gravity-thermodynamics conjecture, predicated on Barrow entropy, is employed in this section to construct the modified Friedmann equations. As we said before, we are considering the apparent horizon as the enveloping horizon. We investigate the application of the generalized second law (GSL) by assuming that the dynamical apparent horizon acts as the thermodynamic boundary. Thermodynamic analysis in holographic cosmological scenarios has been reported in some significant research works including \cite{K3,K4,K5}. To accomplish our objective, we assume that the apparent horizon possesses Barrow entropy.

\begin{figure}
\begin{center}
\includegraphics[height=3.5in]{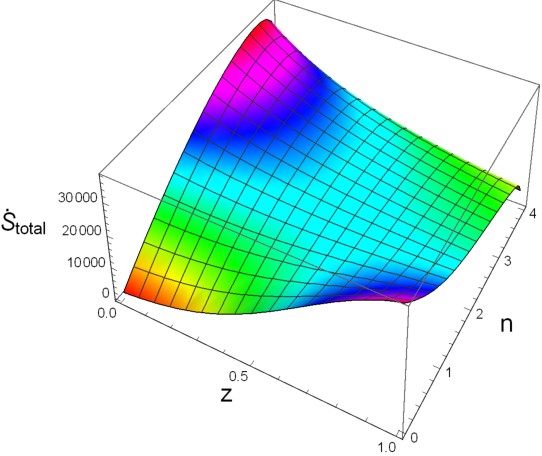}
\caption{\textcolor{black}{Evolution of of the time derivative of total entropy $\dot{S}_{total}$ with respect to redshift $z$ and $n$ for the BHVMCG model. The parameters chosen are $A = 1$, $\gamma = 3.34$, $B_{0}=0.9$, $\alpha=0.04$, $\Delta =0.007$,  $\rho_{VMCG0}=0.006$ and $A_{0}=0.002$.}}
\label{F6}
\end{center}
\end{figure}

We have assumed that $a_{0}=a=1$ at $t=t_{0}$. The Hayward temperature of the
apparent horizon is defined either by \cite{49}
\begin{equation}
T_{h}=-\frac{1}{2\pi \tilde{r}_{a}}\left(1-\frac{\dot{\tilde{r}}_{a}}{2 H \tilde{r}_{a}}\right),
    \label{E28}
\end{equation}
or by \cite{50,51}
\begin{equation}
T_{h}=\frac{1}{2\pi \tilde{r}_{a}}\left(1-\frac{\dot{\tilde{r}}_{a}}{2 H \tilde{r}_{a}}\right),
    \label{E29}
\end{equation}
by following the Hawking temperature. In detail, the abovementioned forms of $T_{h}$ have been discussed in \cite{52}. Here we consider the form Eq.(\ref{E28}) of $T_{h}$. From Eq.(\ref{E10}), we have the apparent horizon $\tilde{r}_{a}$ and its time derivative $\dot{\tilde{r}}_{a}$ as
\begin{equation}
\tilde{r}_{a}=\left(\frac{\left(a^{-3 (1+A) (1+\alpha )} \rho_{VMCG0}+\frac{3 a^{-n} B_{0} (1+\alpha )}{3-n+3 \alpha +3 A (1+\alpha )}\right)^{\frac{1}{1+\alpha
}}}{\gamma}\right)^{\frac{1}{\Delta-2 }},
    \label{E30}
\end{equation}
and
\begin{equation}
\begin{array}{c}
\dot{\tilde{r}}_{a}=a \left(a^{-3 (1+A) (1+\alpha )} \rho_{VMCG0}+\frac{3 a^{-n} B_{0} (1+\alpha )}{3-n+3 \alpha +3 A (1+\alpha )}\right)^{-1+\frac{1}{1+\alpha
}}\times\\
\left(\frac{\left(a^{-3 (1+A) (1+\alpha )} \rho_{VMCG0}+\frac{3 a^{-n} B_{0} (1+\alpha )}{3-n+3 \alpha +3 A (1+\alpha )}\right)^{\frac{1}{1+\alpha
}}}{\gamma}\right)^{-1+\frac{1}{2-\Delta }+\frac{1}{-2+\Delta }} \times\\
\left(-3 a^{-1-3 (1+A) (1+\alpha )} (1+A) \rho_{VMCG0} (1+\alpha )-\frac{3 a^{-1-n} B_{0}
n (1+\alpha )}{3-n+3 \alpha +3 A (1+\alpha )}\right)(\gamma (1+\alpha ) (-2+\Delta ))^{-1}.
\end{array}
\label{E31}
\end{equation}
Using Eqs. (\ref{E10}), (\ref{E30}) and (\ref{E31}) in Eq.(\ref{E28}), we have obtained the reconstructed Hayward temperature of the apparent horizon as
\begin{equation}
\begin{array}{c}
T_{h}=\left(\frac{\left(a^{-3 (1+A) (1+\alpha )} \rho_{VMCG0}+\frac{3 a^{-n} B_{0} (1+\alpha )}{3-n+3 \alpha +3 A (1+\alpha
)}\right)^{\frac{1}{1+\alpha }}}{\gamma}\right)^{\frac{1}{2-\Delta }} \\
\left(-3 a^{3 (1+A) (1+\alpha )} B_{0} (n+2 (1+\alpha ) (-2+\Delta ))+a^n \rho_{VMCG0}
(-3+n-3 \alpha -3 A (1+\alpha )) (-1+3 A+2 \Delta )\right)\\
\left(4 \pi  \left(3 a^{3 (1+A) (1+\alpha )} B_{0} (1+\alpha )+a^n \rho_{VMCG0} (3-n+3 \alpha
+3 A (1+\alpha ))\right) (-2+\Delta )\right)^{-1}.
\end{array}
\label{E32}
\end{equation}
Following \cite{53,54}, we proceed with the entropy computation for the current scenario. To apply the gravity-thermodynamics conjecture, we propose that the first law of thermodynamics on the apparent horizon meets the following conditions:
\begin{equation}
dE=T_{h}dS_{h}+WdV,
    \label{E33}
\end{equation}
where $S_{h}$ and $T_{h}$ are the entropy and temperature associated with the apparent horizon \cite{53} respectively.  $W=\frac{\rho-p}{2}$ \cite{53,55} is the work density connected with the volume change of the expanding universe. In Eq.(\ref{E33}), the total energy of the universe enclosed by the apparent horizon is given by $E=\rho V$. $A=4 \pi \tilde{r}_{a}^{2}$ is the area of the apparent horizon and the volume enclosed by the area of the apparent horizon is given by $V=\frac{4}{3}\pi \tilde{r}_{a}^{3}$. From the relation $E=\rho V$, we have 
\begin{equation}
dE=r\pi \tilde{r}_{a}^{2}\rho d\tilde{r}_{a}-4\pi \tilde{r}_{a}^{3}\dot{\rho}dt.
    \label{E34}
\end{equation}
By the conservation equation, Eq.(\ref{E34}) can be rewritten as
\begin{equation}
\dot{E}=4\pi \tilde{r}_{a}^{2}(\rho \dot{\tilde{r}}_{a}-\tilde{r}_{a}H(\rho+p)).
    \label{E35}
\end{equation}
Here $\dot{E}$ is the time derivative of the total energy of the universe encompassed by the apparent horizon. In our study, dark energy and dark matter for the BHVMCG model have been considered. Eq.(\ref{E35}) can be broken into two parts as
\begin{equation}
\dot{E}_{DE}=4 \pi \tilde{r}_{a}^{2} (\rho_{DE}  \dot{\tilde{r}}_{a}-\tilde{r}_{a} H(\rho_{DE}+ p_{DE})),
    \label{E36}
\end{equation}
\begin{equation}
\dot{E}_{DM}=4 \pi \tilde{r}_{a}^{2} (\rho_{DM} \dot{\tilde{r}}_{a}-\tilde{r}_{a} H\rho_{DM}).
    \label{E37}
\end{equation}
 By differentiating the Barrow entropy, we can obtain the time derivative of horizon entropy as
 \begin{equation}
\dot{S}_{h}=\left(\frac{4 \pi}{A_{0}}\right)^{1+\frac{\Delta}{2}} (2+\Delta)\tilde{r}_{a}^{1+\Delta}\dot{\tilde{r}}_{a}.
    \label{E38}
\end{equation}
The expression of $\dot{E}$ for the BHVMCG dark energy and dark matter are:
\begin{equation}
\begin{array}{c}
\dot{E}_{DE,reconstruct}=-\left(4 \pi  \left(\frac{\left(a^{-3 (1+A) (1+\alpha )} \rho_{VMCG0}+\frac{3 a^{-n} B_{0} (1+\alpha )}{3-n+3 \alpha +3 A (1+\alpha )}\right)^{\frac{1}{1+\alpha
}}}{\gamma}\right)^{\frac{2}{-2+\Delta }}\right.\\
\left(a^{3 (1+A) (1+\alpha )} B_{0} n+a^n (1+A) \rho_{VMCG0} \right.
\left.\left.(3-n+3 \alpha +3 A (1+\alpha ))\right) \right.\\
\left.\left(\frac{a^n\left(a^{3 (1+A) (1+\alpha )} B_{0} (2 n+3 (1+\alpha ) (-2+\Delta )+3 A (1+\alpha ) (-2+\Delta ))+a^n (1+A) \rho_{VMCG0} (3-n+3 \alpha +3 A (1+\alpha
)) \Delta \right)}{B_{0} \left(3 a^{3 (1+A) (1+\alpha )} B_{0} (1+\alpha )+a^n \rho_{VMCG0} (3-n+3 \alpha +3 A (1+\alpha ))\right) (-2+\Delta
)}\right)^{-\frac{1}{1+\alpha }}\right)\\
\left(\left(3 a^{3 (1+A) (1+\alpha )} B_{0} (1+\alpha )+a^n \rho_{VMCG0} (3-n+3 \alpha +3 A (1+\alpha ))\right) (-2+\Delta)\right)^{-1},
\end{array}
\label{E39}
\end{equation}
and
\begin{equation}
\begin{array}{c}
\dot{E}_{DM,reconstruct}=\\
\left(4 \pi  \left(\frac{a^n \left(a^{3 (1+A) (1+\alpha )} B_{0} (2 n+3 (1+\alpha ) (-2+\Delta )+3 A (1+\alpha ) (-2+\Delta ))+a^n
(1+A) \rho_{VMCG0} (3-n+3 \alpha +3 A (1+\alpha )) \Delta \right)}{B_{0} \left(3 a^{3 (1+A) (1+\alpha )} B_{0} (1+\alpha )+a^n \rho_{VMCG0} (3-n+3
\alpha +3 A (1+\alpha ))\right) (-2+\Delta )}\right)^{-\frac{1}{1+\alpha }}\right.\\
\left.\left(3 a^{3 (1+A) (1+\alpha )} B_{0} (n+(1+\alpha ) (-2+\Delta ))+a^n
\rho_{VMCG0} (3-n+3 \alpha +3 A (1+\alpha )) (1+3 A+\Delta )\right)\right. \\
\left.\left(\left(\frac{\left(a^{-3 (1+A) (1+\alpha )} \rho_{VMCG0}+\frac{3 a^{-n} B_{0}
(1+\alpha )}{3-n+3 \alpha +3 A (1+\alpha )}\right)^{\frac{1}{1+\alpha }}}{\gamma}\right)^{\frac{2}{-2+\Delta }}-\right.\right.\\
\left.\left.3 \left(\frac{a^n \left(a^{3 (1+A) (1+\alpha
)} B_{0} (2 n+3 (1+\alpha ) (-2+\Delta )+3 A (1+\alpha ) (-2+\Delta ))+a^n (1+A) \rho_{VMCG0} (3-n+3 \alpha +3 A (1+\alpha )) \Delta \right)}{B_{0}
\left(3 a^{3 (1+A) (1+\alpha )} B_{0} (1+\alpha )+a^n \rho_{VMCG0} (3-n+3 \alpha +3 A (1+\alpha ))\right) (-2+\Delta )}\right)^{\frac{1}{1+\alpha
}}\right)\right)\\
\left(\left(3 a^{3 (1+A) (1+\alpha )} B_{0} (1+\alpha )+a^n \rho_{VMCG0} (3-n+3 \alpha +3 A (1+\alpha ))\right) (-2+\Delta )\right)^{-1},
\end{array}
\label{E40}
\end{equation}
respectively. The time derivative of entropy on the horizon due to BHVMCG dark energy can be obtained by using Eqs. (\ref{E30}) and (\ref{E31}) in Eq.(\ref{E38}) as 
\begin{equation}
\begin{array}{c}
\dot{S}_{h}=-\left(3 \times 2^{2+\Delta } \left(\frac{1}{A_{0}}\right)^{1+\frac{\Delta }{2}} \pi ^{1+\frac{\Delta }{2}} \left(\left(\frac{\left(a^{-3
(1+A) (1+\alpha )} \rho_{VMCG0}+\frac{3 a^{-n} B_{0} (1+\alpha )}{3-n+3 \alpha +3 A (1+\alpha )}\right)^{\frac{1}{1+\alpha }}}{\gamma}\right)^{\frac{1}{-2+\Delta
}}\right)^{1+\Delta }\right.\\
\left.\left(a^{3 (1+A) (1+\alpha )} B_{0} n+a^n (1+A) \rho_{VMCG0} (3-n+3 \alpha +3 A (1+\alpha ))\right) (2+\Delta )\right)\\
\left(\left(3 a^{3
(1+A) (1+\alpha )} B_{0} (1+\alpha )+a^n \rho_{VMCG0} (3-n+3 \alpha +3 A (1+\alpha ))\right) (-2+\Delta )\right)^{-1}.
\end{array}
\label{E41}
\end{equation}
Since a temperature differential \cite{10,56,57} could lead to non-equilibrium thermodynamics, it is typical to assume in this context that the dynamical apparent horizon and the internal fluid temperature in gravitational thermodynamics are equal. Now, the temperature of the fluid inside the horizon will be identified with the temperature on the horizon. We have \cite{10}, from the first law of thermodynamics
\begin{equation}
\dot{S}_{DM}=\frac{1}{T_h}\left(p_{DM}\dot{V}+\dot{E}_{DM}\right),
    \label{E42}
\end{equation}
\begin{equation}
\dot{S}_{DE}=\frac{1}{T_h}\left(p_{DE}\dot{V}+\dot{E}_{DE}\right).
    \label{E43}
\end{equation}
The time derivatives of the dark energy and dark matter inside the horizon for the BHVMCG model are as follows:
\begin{equation}
\begin{array}{c}
\dot{S}_{DM,reconstruct}=\left(16 \pi ^2 \left(\frac{\left(a^{-3 (1+A) (1+\alpha )} \rho_{VMCG0}+\frac{3 a^{-n} B_{0} (1+\alpha )}{3-n+3 \alpha +3 A (1+\alpha
)}\right)^{\frac{1}{1+\alpha }}}{\gamma}\right)^{\frac{1}{-2+\Delta }}\right.\\
 \left(\frac{a^n \left(a^{3 (1+A) (1+\alpha )} B_{0} (2 n+3 (1+\alpha ) (-2+\Delta )+3 A (1+\alpha ) (-2+\Delta ))+a^n (1+A) \rho_{VMCG0} (3-n+3
\alpha +3 A (1+\alpha )) \Delta \right)}{B_{0} \left(3 a^{3 (1+A) (1+\alpha )} B_{0} (1+\alpha )+a^n \rho_{VMCG0} (3-n+3 \alpha +3 A (1+\alpha
))\right) (-2+\Delta )}\right)^{-\frac{1}{1+\alpha }} \\
\left(3 a^{3 (1+A) (1+\alpha )} B_{0} (n+(1+\alpha ) (-2+\Delta ))+a^n \rho_{VMCG0} (3-n+3 \alpha +3 A (1+\alpha )) (1+3 A+\Delta )\right)\\
\left(\left(\frac{\left(a^{-3 (1+A) (1+\alpha )} \rho_{VMCG0}+\frac{3 a^{-n} B_{0} (1+\alpha )}{3-n+3 \alpha +3 A (1+\alpha )}\right)^{\frac{1}{1+\alpha
}}}{\gamma}\right)^{\frac{2}{-2+\Delta }}-\right.\\
\left.\left.3 \left(\frac{a^n \left(a^{3 (1+A) (1+\alpha )} B_{0} (2 n+3 (1+\alpha ) (-2+\Delta )+3 A (1+\alpha ) (-2+\Delta ))+a^n
(1+A) \rho_{VMCG0} (3-n+3 \alpha +3 A (1+\alpha )) \Delta \right)}{B_{0} \left(3 a^{3 (1+A) (1+\alpha )} B_{0} (1+\alpha )+a^n \rho_{VMCG0} (3-n+3
\alpha +3 A (1+\alpha ))\right) (-2+\Delta )}\right)^{\frac{1}{1+\alpha }}\right)\right)\\
\left(-3 a^{3 (1+A) (1+\alpha )} B_{0} (n+2 (1+\alpha ) (-2+\Delta ))+a^n \rho_{VMCG0} (-3+n-3 \alpha -3 A (1+\alpha )) (-1+3 A+2 \Delta )\right)^{-1},
\end{array}
\label{E44}
\end{equation}
\begin{equation}
\begin{array}{c}
\dot{S}_{DE,reconstruct}=-\left(\left(32 \pi ^2 \left(\frac{\left(a^{-3 (1+A) (1+\alpha )} \rho_{VMCG0}+\frac{3 a^{-n} B_{0} (1+\alpha )}{3-n+3 \alpha +3
A (1+\alpha )}\right)^{\frac{1}{1+\alpha }}}{\gamma}\right)^{\frac{3}{-2+\Delta }} \right.\right.\\
\left(a^{3 (1+A) (1+\alpha )} B_{0} n+a^n (1+A) \rho_{VMCG0} (3-n+3\alpha +3 A (1+\alpha ))\right) \\
\left(\frac{a^n \left(a^{3 (1+A) (1+\alpha )} B_{0} (2 n+3 (1+\alpha ) (-2+\Delta )+3 A (1+\alpha ) (-2+\Delta ))+a^n (1+A) \rho_{VMCG0} (3-n+3
\alpha +3 A (1+\alpha )) \Delta \right)}{B_{0} \left(3 a^{3 (1+A) (1+\alpha )} B_{0} (1+\alpha )+a^n \rho_{VMCG0} (3-n+3 \alpha +3 A (1+\alpha
))\right) (-2+\Delta )}\right)^{-\frac{1}{1+\alpha }} \\
\left.\left(3 a^{3 (1+A) (1+\alpha )} B_{0} (n+(1+\alpha ) (-2+\Delta ))+a^n \rho_{VMCG0} (3-n+3 \alpha +3 A (1+\alpha )) (1+3 A+\Delta
)\right)\right)\\
\left(\left(3 a^{3 (1+A) (1+\alpha )} B_{0} (1+\alpha )+a^n \rho_{VMCG0} (3-n+3 \alpha +3 A (1+\alpha ))\right) (-2+\Delta )\right.\\
\left.\left. \left(3 a^{3 (1+A) (1+\alpha )} B_{0} (n+2 (1+\alpha ) (-2+\Delta ))+a^n \rho_{VMCG0} (3-n+3 \alpha +3 A (1+\alpha )) (-1+3 A+2
\Delta )\right)\right)^{-1}\right).
\end{array}
\label{E45}
\end{equation}
\textcolor{black}{By adding Eqs. (\ref{E41}), (\ref{E44}) and (\ref{E45}), we obtain the time derivative of the total entropy $\dot{S}_{total}$ for the BHVMCG model, and the same has been plotted in Fig.\ref{F6}. In this plot, we have varied the redshift $z$ and $n$ to view the behaviour of $\dot{S}_{total}$. Fig.\ref{F6} shows that $\dot{S}_{total}$ remains in the positive level. This shows the validity of the GSL of thermodynamics for the BHVMCG model which means the result is consistent with \cite{59}. If we have a close look into this Fig.\ref{F6}, we observe that for the values of $n$ very close to $0$, $\dot{S}_{total}$ is showing a decreasing pattern with the evolution of the universe. However in this case the $\dot{S}_{total}$ is tending to $0$ with the evolution of the universe. This decaying pattern of $\dot{S}_{total}$ is retained for a very small span of $n$ near $0$. Furthermore at $n=0.2$, despite the decaying pattern it is not tending to $0$ and is significantly above $0$. The surface observed in Fig.\ref{F6} clearly shows that for the current universe $\dot{S}_{total}$ has a sharp increase above $0$ with the increase in the values of $n$. However, it is noteworthy that in the early stage of the universe $\dot{S}_{total}$ is decreasing with $n$. Furthermore, from $n \geq 1.9$ $\dot{S}_{total}$ has a visible increasing pattern with the evolution of the universe. Hence we interpret that although the GSL is verified for this model, the values of the model parameter $n$ significantly impact its behaviour with the evolution of the universe.} 

Let us now discuss the thermodynamic significance of the quintessence to phantom transition. Heat would always flow from the negative-temperature universe into the positive-energy universe, if two copies of the current universe one with a positive temperature and the other with a negative temperature are placed in thermal contact. This would suggests that, at the time a phantom energy becomes dominant i.e. $w_{total,reconstruct} < -1$, an inevitably ``hotter" cosmic evolution regime emerges which is in agreement with \textcolor{black}{\cite{58,T1,T2,T3,T4,T5,T6}}. By considering the effect of the transition from the quintessence to phantom dominated universe on the generalized second law, we may calculate the time derivative of the enclosing horizon as well as the time derivative inside the horizon. The transition happened around $z \approx 1.53$ and the entropy was nearly zero at that time. In the summary, we examined the phantom and quintessence dominated universes and showed that the overall entropy does not decrease with the evolution of the universe. Otherwise, the second law of thermodynamics is rendered invalid. \textcolor{black}{This result is in agreement with \cite{59}, where the author considered the total entropy of the universe as the entropy of the event horizon, normal scalar, and ghost scalar field and shows that phantom entropy must increase with the expansion of the universe. In the next section, we have demonstrated the validity of our model with observational data.} 

\section{Observational Constraints}
In light of the cosmology discussed in the preceding sections, this section serves as a preface to the examining observational data. Let's first discuss the computational basis of the analysis of the observational data before moving on to the analysis described in the following sections. 
There is a dimensionless parameter for the expansion rate which is defined as
\begin{equation}
E(z)=\frac{H(z)}{H_{0}},
\label{O1}
\end{equation}
where $H_{0}$ is the present value of Hubble parameter. From Eq.(\ref{O1}), we can re-express the reconstructed expression of Hubble parameter Eq.(\ref{E10}) in terms of redshift $z$ (where $a=(1+z)^{-1}$) as
\begin{equation}
H(z)=\Omega_{bhd0}\left[(1+z)^{3(1+\alpha)(1+A)}\Omega_{VMCG0}+(1+z)^{n} \Omega_{B0} \right]^{\frac{1}{(1+\alpha)(2-\Delta)}},
\label{O2}
\end{equation}
where $\Omega_{bhd0}=\left(\frac{1}{\gamma}\right)^{\frac{1}{2-\Delta}}(3 H_{0}^{2})^{\frac{1}{(1+\alpha)(2-\Delta)}}$, $\Omega_{VMCG0}=\frac{\rho_{VMCG0}}{3 H_{0}^{2}}$ and $\Omega_{B0}=\frac{B_{0}(1+\alpha)}{H_{0}^{2}[3(1+\alpha)(1+A)-n]}$.

\section{Observational Data Analysis for the model}
In this section, Stern data has been used to conduct a thorough observational data analysis \cite{B1,B2,B3,B4,B5,B05,P1,P2} for the model under consideration. The model has also been examined using joint analyses of Stern$+$BAO and Stern$+$BAO$+$CMB. In the current work, the technique used is the $\chi^{2}$ minimum test of the theoretical Hubble parameter with the observed data set and finding the best-fit values of unknown parameters for various confidence levels (66 $\%$, 90 $\%$, 99 $\%$). 

\textcolor{black}{At this juncture let us have some theoretical discussion on $\chi^2$ minimization \cite{book,book1} and its use in testing goodness of fit. If we consider a set of observations and a model described by a set of parameters $\theta$, it is our job to fit the model to the data. There may be some physical motivation or some function that is convenient to handle. It is then required to define a merit function that quantifies the agreement between the data and the model. This is done by maximizing the agreement that one can obtain through best-fit parameters. Any procedure utilized for fitting requires providing best-fit parameters, estimation of error on the parameters, and a possible measure of the goodness of fit. The details of the methodology are discussed in \cite{book}. The least squares in its simplest form \cite{book} is 
\begin{equation}
    \chi^2 =\sum_{i}w_i[D_i-y(x_i\mid \theta)]^2,
    \label{C1}
\end{equation}}
\textcolor{black}{where $w_i$ are suitably defined weights and $D_i$ is the set of data points. One can show that the minimum variance weight is $w_i=\frac{1}{\sigma_i^2}$, where $\sigma_i$ denotes the error on data point $i$. In this case, the least squares is called chi-square $\chi^2$. If the data are correlated, the chi-square becomes \cite{book}
\begin{equation}
\chi^2 =\sum_{ij}(D_i-y(x_i\mid \theta))Q_{ij}(D_j-y(x_j\mid \theta)),
    \label{C2}
\end{equation}
where $Q$ denotes the inverse of the so called covariance matrix describing the covariance between the data. The best-fit parameters are those parameters that minimize the chi-square $\chi^2$.}

\subsection{Stern $(H(z)-z)$ Data Set}
\begin{figure}
\begin{center}
\includegraphics[height=3.5in]{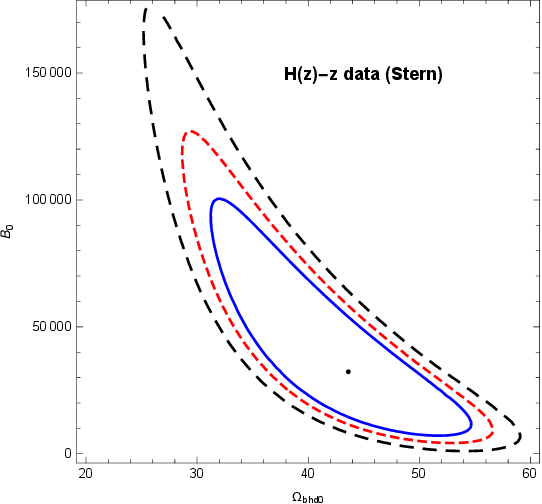}
\caption{The variations of $B_{0}$ against $\Omega_{bhd0}$ for the $(H(z)-z)$ joint analysis. The contours are drawn for 66 $\%$ (solid, blue), 90 $\%$ (dashed, red), and 99 $\%$ (dashed, black) confidence levels. The parameters chosen are $\alpha=0.001$ $\Omega_{VMCG0}=0.75$, $A=0.001$, $\Delta=0.00001$, $\gamma=2.5$, $n=0.001$, and $H_{0}=72$.}
\label{F8}
\end{center}
\end{figure}

The BHVMCG model has been examined in this subsection using the observed Hubble parameter values for the various redshifts (12 data points) mentioned by Stern et al. \cite{B6} in the observed Hubble data set.  In Table \ref{T1}, we have the observed values of Hubble parameter $H(z)$ and the standard error $\sigma(z)$ for different values of redshift $z$. The validity of a statistical hypothesis is assessed at various confidence levels. 
\begin{table}[h!]
\caption{The Hubble parameter $H(z)$ and the standard error $\sigma(z)$ are shown tabularly for a range of redshift $z$ values \cite{B6}.}
\centering
 \begin{tabular}{||c | c | c||} 
 \hline
~~~~~$z$~~~~~ & ~~~~~$H(z)$~~~~~ & ~~~~~$\sigma(z)$~~~~~\\ [1ex] 
 \hline\hline
 0 & 73 & $\pm8$ \\[1ex]
 \hline
 0.1 & 69 & $\pm12$ \\[1ex]
 \hline
 0.17 & 83 & $\pm8$ \\[1ex]
 \hline
 0.27 & 77 & $\pm14$ \\[1ex]
 \hline
 0.4 & 95 & $\pm17.4$ \\[1ex]
 \hline
 0.48 & 90 & $\pm60$ \\[1ex]
 \hline
 0.88 & 97 & $\pm40.4$ \\[1ex]
 \hline
 0.9 & 117 & $\pm23$ \\[1ex]
 \hline
 1.3 & 168 & $\pm17.4$ \\[1ex]
 \hline
 1.43 & 177 & $\pm18.2$ \\[1ex]
 \hline
 1.53 & 140 & $\pm14$ \\[1ex]
 \hline
 1.75 & 202 & $\pm40.4$ \\[1ex]
 \hline\hline
 \end{tabular}
 \label{T1}
\end{table}
To achieve this, the $\chi^{2}$ statistics is first formed as a sum of the standard normal distribution as follows:
\begin{equation}
\chi^{2}_{Stern}=\sum\frac{(H(z)-H_{obs}(z))^{2}}{\sigma^
{2}(z)}
\label{O3}
\end{equation}
\begin{equation}
L=\int e^{-\frac{1}{2}\chi^{2}_{Stern}}P(H_{0})dH_{0},
    \label{O4}
\end{equation}
where $H_{obs}(z)$ and $H(z)$ are observational and theoretical values of the Hubble parameter for various values of redshifts respectively. $H_{obs}(z)$ can be securely marginalized since it is a nuisance parameter. The current value of Hubble parameter is $H_{0} = 72 \pm 8$ Km$s^{-1}$ Mp$c^{-1}$. The best-fit value of the parameters ($B_{0}$ vs  $\Omega_{bhd0}$) can be determined by minimizing the abovementioned distribution $\chi^{2}_{Stern}$ and the other unknown parameters can be fixed with the help of Stern data. According to our analysis, the best-fit values of $B_{0}$ against  $\Omega_{bhd0}$ are presented in Table \ref{T2}. The graph for different confidence levels is plotted. The theoretical range of the parameters is supported by our best-fit analysis with Stern observational data. In Fig.\ref{F8}, we have plotted the contours of $B_{0}$ vs $\Omega_{bhd0}$ for 66 $\%$
(solid, blue), 90 $\%$ (dashed, red) and 99 $\%$ (dashed, black) confidence levels for the BHVMCG model.
\begin{table}[h!]
\caption{The best-fit values of $B_{0}$,
$\Omega_{bhd0}$ and the minimum values of $\chi^{2}$}
\centering
 \begin{tabular}{||c | c | c | c ||} 
 \hline
~~~~~~Data~~~~~~ & ~~~~~~$B_{0}$~~~~~~ & ~~~~~~$\Omega_{bhd0}$~~~~~~&~~~~~~$\chi^{2}_{min}$~~~~~~\\ [1ex] 
 \hline\hline
 Stern & 32315.8 & 43.6103 & 7.34303 \\[1ex]
 \hline
 Stern+BAO & 64550.8 & 30.8480 & 768.307 \\[1ex]
 \hline
 Stern+BAO+CMB & 66926.9 & 30.3193 & 9962.83 \\[1ex]
 \hline\hline
 \end{tabular}
 \label{T2}
\end{table}

\subsection{Stern $+$ BAO Data Sets}
\begin{figure}
\begin{center}
\includegraphics[height=3.5in]{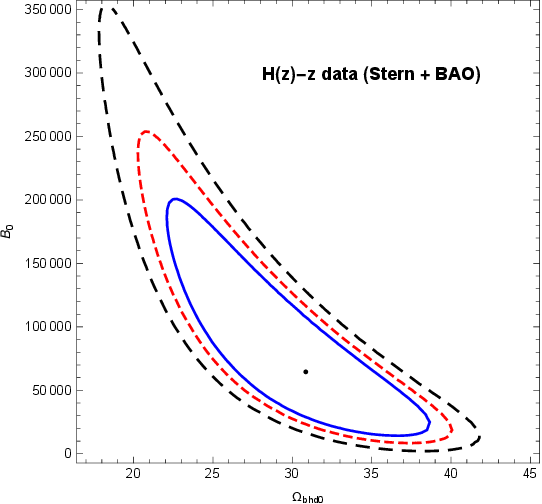}
\caption{The variations of $B_{0}$ against $\Omega_{bhd0}$ for the $((H(z)-z)+ BAO)$ joint analysis. The contours are drawn for 66 $\%$ (solid, blue), 90 $\%$ (dashed, red), and 99 $\%$ (dashed, black) confidence levels. The parameters chosen are $\alpha=0.001$ $\Omega_{VMCG0}=1.5$, $A=0.001$, $\Delta=0.00001$, $\gamma=2.5$, $n=0.001$, $\Omega_{m}=0.01$, and $H_{0}=72$.}
\label{F9}
\end{center}
\end{figure}

In this subsection, a joint analysis is used in the sense that the Baryon Acoustic Oscillation (BAO) peaks are included in the stern data. The BAO peak parameter value was proposed by Eisenstein et al. \cite{B7}  and their methodology has been used here. As far as the detection of BAO signal
is concerned, the Sloan Digital Sky Survey (SDSS) is considered the pioneer. The BAO signals were immediately identified by the survey at a scale of about 100 MPc. The combination of angular diameter distance and Hubble parameter at that redshift is the analysis that is performed. This analysis does not include any specific dark energy and is not dependent on the measurement of $H_{0}$. In this study, we have used standard $chi^{2}$ analysis to investigate the parameters $B_{0}$ vs $\Omega_{bhd0}$ for the BHVMCG model from the BAO peak data for low redshift ($0<z<0.35$). In the case of a Gaussian distribution, the error is equivalent to the standard deviation. We know that different cosmological parameters like the EoS parameter of dark energy have less influence on Low-redshift distance measurements and can directly measure the Hubble constant $H_{0}$. The BAO peak parameter is defined as 
\begin{equation}
    \mathcal{A}=\frac{\sqrt{\Omega_{m}}}{E(z_{1})^{\frac{1}{3}}}\biggl(\frac{1}{z_{1}}\int_{0}^{z_{1}}\frac{dz}{E(z)}\biggr)^{\frac{2}{3}},
    \label{O5}
\end{equation}
where the normalized Hubble parameter is $E(z)$, the usual redshift of the SDSS sample is $z_{1}=0.35$, and the integration term is the dimensionless comoving distance at the redshift $z_{1}$. Using SDSS data \cite{B7} from the luminous red galaxies survey, the value of the parameter $\mathcal{A}$ is given by $\mathcal{A}=0.469 \pm 0.017$ for the flat model of the universe. For the BAO measurement, the $\chi^{2}$ function is 
\begin{equation}
    \chi^{2}_{BAO}=\frac{(\mathcal{A}-0.469)^{2}}{(0.017)^{2}}.
    \label{O6}
\end{equation}
The total joint analysis of (Stern $+$ BAO) data sets for the $\chi^{2}$ function is defined by
\begin{equation}
    \chi^{2}_{Total}= \chi^{2}_{Stern}+ \chi^{2}_{BAO}.
    \label{O7}
\end{equation}
According to our analysis, the best-fit values of $B_{0}$ vs $\Omega_{bhd0}$ for the joint data analysis of $(H(z)-z+ BAO)$ are presented in Table \ref{T2}. In Fig.\ref{F9}, we have generated the closed contours of $B_{0}$ vs $\Omega_{bhd0}$  for 66 $\%$ (solid, blue), 90 $\%$ (dashed, red) and 99 $\%$ (dashed, black) confidence levels for the BHVMCG model.

\subsection{Stern $+$ BAO $+$ CMB Data Sets}
\begin{figure}
\begin{center}
\includegraphics[height=3.5in]{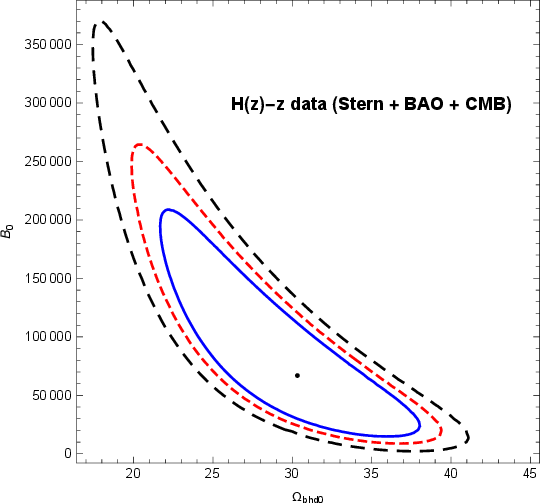}
\caption{The variations of $B_{0}$ against $\Omega_{bhd0}$ for the $((H(z)-z)+ BAO + CMB)$ joint analysis.The contours are drawn for 66 $\%$ (solid, blue), 90 $\%$ (dashed, red), and 99 $\%$ (dashed, black) confidence levels. The parameters chosen are $\alpha=0.001$ $\Omega_{VMCG0}=1.55$, $A=0.001$, $\Delta=0.00001$, $\gamma=2.5$, $n=0.001$, $\Omega_{m}=0.01$, and $H_{0}=72$.}
\label{F10}
\end{center}
\end{figure}

The angular scale of the sound horizon at the surface of the last scattering measures the angular scale of the first acoustic peak. This is regarded as one of the most intriguing geometrical probes of dark energy. The CMB (Cosmic Microwave Background) power spectrum encodes the information. The definition of the CMB shift parameter is defined in \cite{B8,B9,B10}. It is not sensitive to perturbations but is suitable to confine the model parameters. In our analysis, we have used this property. The shift parameter is the CMB power spectrum's first peak which is given by
\begin{equation}
    \mathcal{R}=\sqrt{\Omega_{m}}\int^{z_{2}}_{0}\frac{dz}{E(z)},
    \label{O8}
\end{equation}
where $z_{2}$ corresponds to the redshift value at the surface of the last scattering. Based on WMAP7 data of Komatsu et al.'s work \cite{B11}, the shift parameter's value was obtained as $\mathcal{R}=1.726 \pm 0.018$ at redshift $z=1091.3$. The $\chi^{2}$ function for the measurement of CMB can be expressed as
\begin{equation}
    \chi^{2}_{CMB}=\frac{(\mathcal{R}-1.726)^{2}}{(0.018)^{2}}.
    \label{O9}
\end{equation}
Hence, when three cosmological tests are considered together, the total joint analysis of data sets (Stern $+$ BAO $+$ CMB) for the $\chi^{2}$ function is given as
\begin{equation}
    \chi^{2}_{Total}= \chi^{2}_{Stern} + \chi^{2}_{BAO} + \chi^{2}_{CMB} .
    \label{O10}
\end{equation}
According to our analysis, the best-fit values of $B_{0}$ vs $\Omega_{bhd0}$ for the joint data analysis of $(Stern $+$ BAO $+$ CMB)$ are presented in Table \ref{T2}, which support the theoretical range of the parameters. The closed contours of $B_{0}$ vs $\Omega_{bhd0}$  for 66 $\%$ (solid, blue), 90 $\%$ (dashed, red) and 99 $\%$ (dashed, black) confidence levels are plotted in Fig.\ref{F10} for the BHVMCG model..

\subsection{Supernovae Type Ia: Redshift-Magnitude Observations}
\begin{figure}
\begin{center}
\includegraphics[height=3.5in]{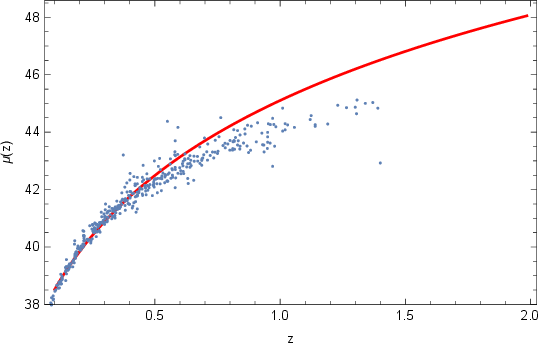}
\caption{$\mu(z)$ vs $z$ curve. The parameters chosen are $\alpha=0.001$ $\Omega_{VMCG0}=0.75$, $A=0.001$, $\Delta=0.00001$, $\gamma=2.5$, $n=0.001$, $\Omega_{bhd0}=30$, $B_{0}=0.1$ and $H_{0}=72$.}
\label{F11}
\end{center}
\end{figure}

Supernova Type Ia experiments provided the main evidence for the existence of dark energy. The existence of dark energy directly corresponds to the redshift of the universe. Since 1995, the Supernova Cosmology Project and the High-z Supernova Search Team have been extensively working, and as a result of their efforts they have discovered various type Ia Supernovae at high redshifts \cite{obs2,B11,B12,B13}. The distance modulus of a Supernovae and its redshift $z$ \cite{B12,B14} is directly measured by the observations. The current observational data have been considered here, including SNe Ia which consists of 557 data points and part of the Union2 sample \cite{B15}. From the observations, the dark energy density is determined by the luminosity distance $d_{L}(z)$ which is defined by
\begin{equation}
    d_{L}(z)=(1+z)H_{0}\int^{z}_{0} \frac{dz^{'}}{H(z^{'})}.
    \label{M11}
\end{equation}
The distance between the absolute and apparent luminosity of a distant object is known as the distance modulus for Supernovae which is given by
\begin{equation}
    \mu(z)=5 log_{10}\biggl[\frac{d_{L}(z)/ H_{0}}{1 MPc}\biggr]+25.
    \label{M12}
\end{equation}
\textcolor{black}{In Fig.\ref{F11}, we have plotted the best-fit of distance modulus $\mu(z)$ as a function of redshift $z$ for our theoretical model (as depicted as a curve in red), and the Supernova Type Ia Union2 sample (as depicted by the dots in blue) considering the best-fit values of the parameters. From the curve, we can conclude that the BHVMCG model is in good agreement with the union2 sample data.}

\section{Concluding remarks}
In this work, we have considered the FRW universe and reconstructed the variable modified Chaplygin gas scenario in the Barrow holographic dark energy framework. Our goal is to rebuild the Hubble parameter to produce Barrow holographic variable modified Chaplygin gas dark energy. The Barrow holographic dark energy, which is characterized by the most generalized cut-off \cite{C1,C2,C3,C4}, is one particular example of Nojiri-Odintsov holographic dark energy. The variable modified Chaplygin gas is rebuilt and the horizon entropy is determined with the thermodynamic implications of the Barrow entropy \cite{10}, using the apparent horizon as the encompassing horizon of the universe. 

The entropies for both sectors have been computed, considering non-interacting dark energy and dark
matter, and finally, the temporal derivative of the overall entropy has been obtained. Considering the correspondence
between Barrow holographic dark energy and variable modified gas, the Hubble parameter has been reconstructed. With the reconstructed Hubble parameter, we have the reconstructed density for the BHVMCG model. \textcolor{black}{Accordingly, the reconstructed EoS parameter $w_{reconstruct}$ for the BHVMCG model is obtained in Eq.(\ref{E11}) and has been plotted against redshift $z$ and $\Delta$ in Fig.\ref{F1}. We infer from the plot that it shows a quintessence behaviour i.e. $-1<w_{reconstruct}<-\frac{1}{3}$, and the reconstructed EoS parameter exhibits a decaying pattern with the evolution of the universe. In this pictorial presentation, a significant variation of the EoS parameter is observed while varying both  $z$ and $\Delta$. The decreasing pattern of the EoS parameter is always there irrespective of the values of $\Delta$ close to $0$ or $2$. However, it is noteworthy that the EoS parameter is decreasing sharply for higher values of $\Delta$ than those of $\Delta$ close to $0$. However, despite this decaying pattern, the value of the reconstructed EoS parameter is close to $-1$ for the current universe i.e. at $z=0$. In all the cases, it becomes asymptotic in the neighbourhood of phantom boundary $-1$ in the later stage. Thus we understand that for the range of the values of the deformation parameter between $0$ and $2$, the reconstructed EoS parameter behaves like quintessence. It is notable that at $z=0$, $w_{reconstruct} \approx -1$ i.e. it behaves as a cosmological constant. We have also obtained the reconstructed total EoS parameter $w_{total,reconstruct}$ (Eq.(\ref{E15})) and has been plotted against redshift $z$ and $\Delta$ in Fig.\ref{F2}. We infer from the plot that some significant variation is seen in the behaviour of this EoS parameter with the values of the deformation parameter $\Delta$. The EoS parameter is asymptotic in the neighbourhood of $-1$ for the values of $\Delta$ close to $0$, and for such values of $\Delta$, the behaviour of the EoS parameter is quintessence. However, as $\Delta$ goes away from $0$, a significant change becomes visible. The EoS parameter starts exhibiting a quintom behaviour nearly from $\Delta \approxeq 0.1$ i.e. a transition from $w_{total,reconstruct}>-1$ to $w_{total,reconstruct}<-1$ which means there is a crossing of phantom boundary and is monotonously decreasing with the evolution of the universe. It should be noted that for values of $\Delta \gtrapprox 0.1$, the $w_{total,reconstruct}$ exhibits a transition from quintessence to phantom, in contrast to the $w_{reconstruct}$, where quintessence behaviour is witnessed. Furthermore, it should be highlighted that in the later stage of the universe, there is no future escape from phantom even when the $w_{total,reconstruct}$ transits to quintom. Therefore, it suggests that big-rip singularity cannot be avoided for the BHVMCG model.} From the plot of reconstructed deceleration parameter $q_{reconstruct}$ (Eq.(\ref{E21})) against redshift $z$ for the BHVMCG model in Fig.\ref{F3}, we observe that at the very early stage of the universe, $q_{reconstruct} > 0$, roughly around $z > 0.5$,  i.e. the decelerated expansion phase of the universe. A transition is seen in the case of the deceleration parameter $q_{reconstruct}$ at $z \approx 0.5$  from a positive to a negative region, which means the universe gradually transits from the decelerated $(q_{reconstruct} > 0)$ to the accelerated expansion phase $(q_{reconstruct} < 0)$ and the transition occurs at $z_{t} \approx 0.45$ which is consistent with the observations in the literature \cite{12}. In the current stage of the Universe it converges towards -$1$ and becomes asymptotic in its neighborhood of -$1$. Hence, in the case of the BHVMCG model, we infer that a transition is possible from the decelerated expansion phase to the accelerated expansion phase of the universe.

\textcolor{black}{Based on Eqs. (\ref{E24}) and (\ref{E25}), the evolution of the reconstructed Statefinder pair $(r,s)$ for the BHVMCG model has been plotted in Fig.\ref{F4}. From the plot we can conclude that for this model the trajectory of the Statefinder pair starts its evolution from the region $(s>0,r<1)$ and with the evolution of the universe it traverses through the $\Lambda$CDM fixed point $(s=0,r=1)$. As described, we can say that the statefinder trajectory traverses the quintessence phase i.e. $(s>0,r<1)$  to reach $\Lambda$CDM fixed point. At the later stage of the universe, it lies in the Chaplygin gas region i.e. $(s<0,r>1)$. As the trajectory passes through the $\Lambda$CDM fixed point, it confirms strongly that our model circulates the $\Lambda$CDM phase of the universe. Moreover, from the plot, it is also understandable that $s$ tends to go to $-\infty$ with finite $r$. Thus the BHVMCG model also indicates the possibility of interpolation between universe's dust and $\Lambda$CDM phase.} We have also plotted the reconstructed $O_{m}(z)$ diagnostic against redshift $z$ for the BHVMCG model in Fig.\ref{F5} based on Eq.(\ref{E27}). We know that the negative curvature of $O_{m}(z)$ diagnostic trajectories displays the quintessence behaviour of DE, whereas the phantom behaviour corresponds to its positive curvature. The trajectories of the reconstructed $O_{m}(z)$ diagnostic for the BHVMCG model show both regions. 

In the next phase, we studied the thermodynamics of the BHVMCG model. The time derivatives of the entropy on the horizon and the fluid within the horizon have been calculated considering the apparent horizon as the enveloping horizon of the Universe. The sum of time derivatives of the entropies of dark matter and dark energy within
the horizon and the time derivative of the entropy on the horizon leads to the time derivative of the total
entropy $\dot{S}_{total}$. \textcolor{black}{The time derivative of the total entropy $\dot{S}_{total}$ for the BHVMCG model is obtained by adding Eqs. (\ref{E41}), (\ref{E44}) and (\ref{E45}), and the same has been plotted in Fig.\ref{F6}. In this plot, the redshift $z$ and the parameter $n$ are varied to view the behaviour of $\dot{S}_{total}$. The plot shows that $\dot{S}_{total}$ remains in the positive level. This shows that the GSL of thermodynamics holds for the BHVMCG model which means the result is consistent with \cite{59}. If we have a close look into this Fig.\ref{F6}, we observe that $\dot{S}_{total}$ is showing a decreasing pattern with the evolution of the universe for the values of $n$ very close to $0$. However in this case the $\dot{S}_{total}$ is tending to $0$ with the evolution of the universe. This decaying pattern of $\dot{S}_{total}$ is retained for a very small span of $n$ near $0$. Furthermore, despite the decaying pattern, at $n=0.2$ it is not tending to $0$ and is significantly above $0$. The surface observed in the plot clearly shows that for the current universe i.e. at $z=0$, $\dot{S}_{total}$ has a sharp increase above $0$ with the increase in the values of $n$. However, it is notable that in the early stage of the universe $\dot{S}_{total}$ is decreasing with the parameter $n$. Furthermore, $\dot{S}_{total}$ has a visible increasing pattern with the evolution of the universe from $n \geq 1.9$. Hence we can interpret that although the GSL is verified for the BHVMCG model, the values of the model parameter $n$ significantly impact its behaviour with the evolution of the universe.}  

We have presented the Hubble parameter $H$ in terms of the observable parameters $\Omega_{bhd0}$, $\Omega_{VMCG0}$, $\Omega_{B0}$, $H_{0}$ along with the redshift $z$ and other parameters such as $\alpha$, $A$, $n$, $\Delta$, $\gamma$, $\rho_{VMCG0}$ and $B_{0}$. We have selected certain numerical values for these parameters that align with observations. By minimizing the $\chi^{2}$ test, we have determined the boundaries of the arbitrary parameters from the Stern data set \cite{B6}. Next, the best-fit values and the bounds of the parameters $(B_{0},\Omega_{bhd0})$ have also been found as a result of the joint analysis of BAO and CMB observations. The statistical confidence contour of $(B_{0}, \Omega_{bhd0})$ have been plotted for 66 $\%$ (solid, blue), 90 $\%$ (dashed, red) and 99 $\%$ (dashed, black) confidence levels by fixing observable parameters such as $\Omega_{bhd0}$, $\Omega_{VMCG0}$, $\Omega_{B0}$ and $H_{0}$ some other parameters like $\alpha$, $B_{0}$ etc. for Stern, Stern$+$BAO and Stern$+$BAO$+$CMB joint data analysis.

The best-fit values and bounds of the parameters $(B_{0}, \Omega_{bhd0})$ are obtained from the Stern data. In the first row of Table \ref{T2}, we have shown the results and the statistical confidence contour of $(B_{0}, \Omega_{bhd0})$ have been plotted in Fig.\ref{F8} for 66 $\%$ (solid, blue), 90 $\%$ (dashed, red) and 99 $\%$ (dashed, black) confidence levels for the BHVMCG model. We have also obtained the best-fit values and bounds of the parameters $(B_{0}, \Omega_{bhd0})$ for the joint analysis of Stern $+$ BAO data. The output values are displayed in the second row of Table \ref{T2} and plotted in Fig.\ref{F9}, the statistical confidence contour of $(B_{0}, \Omega_{bhd0})$ for 66 $\%$ (solid, blue), 90 $\%$ (dashed, red) and 99 $\%$ (dashed, black) confidence levels for the BHVMCG model. Next, the best-fit values and bounds of the parameters $(B_{0}, \Omega_{bhd0})$ are found for the joint analysis of Stern $+$ BAO $+$ CMB data and the results are shown in the third row of Table \ref{T2} and in Fig.\ref{F10}, the statistical confidence contour of $(B_{0}, \Omega_{bhd0})$ for 66 $\%$ (solid, blue), 90 $\%$ (dashed, red) and 99 $\%$ (dashed, black) confidence levels have been plotted for the BHVMCG model. In each case, the model parameters are compared by their output values and the statistical contours. The comparison analysis provides insight into how the theoretical values of the parameters converge with the values derived from the observational data set and how this varies for different selected sets of other parametric values.

\textcolor{black}{Finally, in Fig.\ref{F11}, we have plotted the best-fit of distance modulus $\mu(z)$ against redshift $z$ for our theoretical model (as depicted as a curve in red), and the  Supernova Type Ia Union2 data sample (as depicted by the dots in blue) considering the best-fit values of the parameters. We have found from the plot that the observational data sets are consistent with our predicted theoretical BHVMCG model.} 

While concluding, let us comment on the overall outcome of the work. We intended to have a phenomenological DE model in the form of variable modified Chaplygin gas in the holographic framework. For this purpose, we have adopted a holographic model through a recently proposed version of it called the Barrow holographic model. The holographically reconstructed Chaplygin dark fluid was found to be capable of attaining the $\Lambda$CDM fixed point and also could go beyond it. The model was assessed by analyzing observational data sets through the $\chi^{2}_{min}$ test and the values obtained through the analyses were observed to be within the region of acceptance proving the goodness of fit of the reconstructed model demonstrated here. Alongside this, a thermodynamic analysis was carried out and the positive time derivative of total entropy proves the consistency of the reconstructed model with the late time acceleration of the universe. We conclude with the future direction of having rigorous analysis of such reconstruction approach in view of the realization of inflationary expansion and thereby unification of early inflation with late time acceleration.

\section{Acknowledgement}
\textcolor{black}{The authors sincerely acknowledge the insightful comments from the anonymous reviewers.} The authors are thankful for the warm hospitality of IUCAA, Pune, India, where a part of the work was carried out in January 2024.

\end{document}